\pdfoutput=1

\documentclass[preprint,fleqn,5p,numbers,sort&compress]{elsarticle}

\usepackage{graphicx}
\usepackage{amssymb,amsmath}
\usepackage{natbib}
\usepackage{hyperref}
\hypersetup{pdftitle = The title of my PDF, pdfauthor = My name, pdfsubject= The subject, pdfkeywords = keyword1 keyword2 keyword3}
\hypersetup{colorlinks = true, linkcolor = blue, anchorcolor = red, citecolor = blue, filecolor = red, pagecolor = red, urlcolor = blue}

\journal{International Journal of Non-Linear Mechanics}

\newcommand{\mathz}{\ooalign{$z$\cr\hfil\rule[.5ex]{.2em}{.06ex}\hfil\cr}}

\begin{document}

\begin{frontmatter}

\title{Revealing the Newton-Raphson basins of convergence in the circular pseudo-Newtonian Sitnikov problem}

\author[eez]{Euaggelos E. Zotos\corref{cor1}}
\ead{evzotos@physics.auth.gr}

\author[mss]{Md Sanam Suraj}

\author[mss]{Mamta Jain}

\author[ra]{Rajiv Aggarwal}

\cortext[cor1]{Corresponding author}

\address[eez]{Department of Physics, School of Science,
Aristotle University of Thessaloniki, GR-541 24, Thessaloniki, Greece}

\address[mss]{Department of Mathematics, Sri Aurobindo College,
University of Delhi, 110017, Delhi, India}

\address[ra]{Department of Mathematics, Deshbandhu College,
University of Delhi, Kalkaji, New Delhi-110019, Delhi, India}

\begin{abstract}
In this paper we numerically explore the convergence properties of the pseudo-Newtonian circular restricted problem of three and four primary bodies. The classical Newton-Raphson iterative scheme is used for revealing the basins of convergence and their respective fractal basin boundaries on the complex plane. A thorough and systematic analysis is conducted in an attempt to determine the influence of the transition parameter on the convergence properties of the system. Additionally, the roots (numerical attractors) of the system and the basin entropy of the convergence diagrams are monitored as a function of the transition parameter, thus allowing us to extract useful conclusions. The probability distributions, as well as the distributions of the required number of iterations are also correlated with the corresponding basins of convergence.
\end{abstract}

\end{frontmatter}

\section{Introduction}
\label{intro}

The Sitnikov problem is a special formulation of the restricted three-body problem which describes the motion of a test particle which oscillates along the vertical $z$-axis, perpendicular to the configuration $(x,y)$ plane, where two primaries, of equal masses, move in circular or elliptic orbits with common barycentre i.e. the axes origin. In particular, when the primaries move in circular orbits, the problem describes the so called MacMillan problem \cite{McM11}. The history of the Sitnikov problem, when the primaries move in circular orbit, begins with Ref. \cite{P07}, where this dynamical model was discussed for the first time. Later on, the Sitnikov problem, of two primary bodies, has been investigated by many authors (see e.g. Refs. \cite{DS97,H09,JE01,JP97,Per07,PM87,PK12,S65,Z17b}).

Several types of perturbations, such as primaries with either prolate \cite{DKMP12} or oblate shape \cite{RGH15}, as well as the radiation pressure \cite{PK06}, have been added for making the system of three bodies more realistic. Another well-studied aspect of the Sitnikov three-body problem is the study of the families of periodic orbits and the corresponding bifurcations \cite{BLO94,KPP08,Per07}. In addition, the stability of motion in the same system has been investigated in \cite{SBV07}, where it was found that in the case where the mass of the test particle is not negligible the energetically allowed regions of motion grow with increasing value of the third body, while at the same time the amplitude of the oscillation, along the vertical $z$-axis, gradually increases.

As the extension of the restricted three-body problem to the restricted four-body is natural, in the same vein the extension of the Sitnikov three-body problem to the Sitnikov four-body problem is quite obvious. In Ref. \cite{SPB08} the existence, the linear stability as well as the bifurcations in the Sitnikov family of straight line periodic orbits in the restricted four-body problem were studied. In this work, it was revealed that the Sitnikov family has only one stability interval, while only twelve 3-dimensional families of symmetric periodic orbits exist, which bifurcate from twelve corresponding critical Sitnikov periodic orbits. On the other hand, the Sitnikov family has infinitely many stability intervals in the restricted three-body Sitnikov problem, which results to infinitely many Sitnikov critical orbits.

Over the years, the Sitnikov four-body problem has been investigated by many authors (see e.g., Ref. \cite{SH11}), including also the oblateness of the primaries (see e.g., Refs. \cite{PA13a,SH13}), the effects of the radiation pressure (see e.g., Ref. \cite{SH14}), and families of periodic orbits and bifurcations (see e.g., Ref. \cite{SPB08}). Moreover, the region of motion in the Sitnikov four-body problem when the fourth mass is finite has also been determined \cite{PA13b}, by exploring the structure of the phase space with the use of proper selection of surfaces of section. It was observed that for low energy, the central fixed point is stable, while for higher values of the energy it splits into an unstable and two stable fixed points.

In Ref. \cite{BP09} the stability of the vertical motion and its bifurcations into families of 3-dimensional periodic orbits in the Sitnikov restricted $N-$body problem were revealed. Additionally, it was found that for $N \geq 4$ there exists only one interval of stable vertical solutions for every $N$. However, as $N$ increases, the stability intervals increase in size. They also demonstrated that a sizable region of bounded orbits close to the $z$-axis also exists. More precisely, these orbits are small near the endpoints of the interval, while they increase in size at the middle of the same interval.

For modelling the motion of the infinitesimal mass (test particle) in the classical $N$-body problem, especially for the three and four body-problems, various, more realistic, modifications have been proposed, mainly by introducing additional perturbative terms to the classical Newtonian effective potential. By using the Einstein-Infeld-Hoffmann theory \cite{EIH38} the first-order post-Newtonian equations of motion, in the restricted three-body problem, have been deduced by several authors (see e.g., Refs. \cite{B72,C76,K67}). Similarly, the orbital dynamics of the planar circular restricted three-body problem, in the context of the pseudo-Newtonian approximation by using the Fodor-Hoenselaers-Perj\'{e} procedure, has been analyzed in Ref. \cite{DLCG17}. Furthermore, in Ref. \cite{HW14} the influence of the separation between the primary bodies was discussed and it was revealed that the post-Newtonian dynamics substantially differ from the corresponding classical Newtonian dynamics, provided that the distances between the primary bodies is sufficiently small.

Undoubtedly, in a dynamical system an issue of great importance is knowing the locations of the equilibrium points. However, in most of the cases, such as in the restricted $N$-body problem (with $N \geq 3$), the position of the libration points cannot be computed by means of analytical methods. Consequently, the only viable alternative is the use of numerical methods. Unfortunately, the outcomes of any numerical technique are directly related to the initial conditions, used to the iterative procedure. More precisely, for initial conditions inside the basins of convergence the iterative procedure lead very quickly to a solution. On the other hand, for initial conditions located in the vicinity of the fractal basin boundaries the numerical scheme usually requires a considerable amount of iterations for reaching to a libration point (root). Therefore, the convergence properties of a system is a highly important issue, because this information reveals the optimal (regarding fast convergence) initial conditions, for numerically locating an equilibrium point.

Over the past decades, the basins of convergence, associated with the libration points, have been investigated mainly by using the multivariate version of the Newton-Raphson method. At the same time, various types of perturbing terms have been added the effective potential of the restricted problem of three and four bodies (see e.g., Refs. \cite{DKMP12,SAA17,SAP17,Z16,Z17a,ZSAS18,Z18b}). Very recently, in Ref. \cite{Z17a} we discussed the basins of convergence, associated with the equilibrium points in the pseudo-Newtonian planar circular restricted three-body problem, by using the multivariate version of the Newton-Raphson iterative scheme. Our analysis revealed that the transition parameter strongly influences the existence and the stability of the libration points as well as the topology of the basins of convergence.

For almost all the $N$-body systems (with $N \geq 3$) we have no analytic equations, regarding the position of the equilibrium points of the systems. This directly implies that we have to use numerical methods for determining the position of the libration points. However, the outcomes of the numerical algorithms are affected by the specific initial conditions, which we use as starting points of the iterative schemes. More precisely, there are initial conditions which display a very fast convergence, while there are also starting points for which the iterative procedure requires a substantial number of iterations before reaching to a solution. The first type of initial conditions compose the so-called ``basins of convergence", while the latter type usually appears in the vicinity of the fractal basin boundaries. Therefore, it is undeniable that knowing the convergence regions of a dynamical system is an issue of paramount importance, because they indicate the optimal initial conditions for the iterative schemes. At the same time, the basins of convergence show us all the fractal regions in which all the corresponding initial conditions should be avoided as starting points.

In Ref. \cite{SL06} it was proved that some pseudo-Newtonian systems are more dynamically stable, compared to their Newtonian equivalent. On this basis, it is very interesting to explore the pseudo-Newtonian version of the Sitnikov problem of three and four primary bodies. The main aim of this work is to unveil the convergence properties of the system by determining how the transition parameter influences the geometry as well as the shape of the Newton-Raphson basins of convergence.

It should be emphasized that the configuration of the circular restricted four-body problem, with equally massed primaries, is dynamically unstable (the corresponding proof is given in \ref{appex}). Nevertheless, we decided to conduct numerical calculations in this system, mainly for comparison reasons of the applied numerical method's behavior to the computation of the equilibrium points, as well as the associated basins of convergence. Additionally, several previous works have been devoted on the four-body problem with three equal masses (see e.g., Refs. \cite{BP11a,BP11b,PP13}).

The present paper has the following structure: the most important properties of the dynamical system are presented in Section \ref{mod}. The parametric evolution of the position of the equilibrium points is investigated in Section \ref{param}. The following Section contains the main numerical results, regarding the evolution of the Newton-Raphson basins of convergence. In Section \ref{bas} we demonstrate how the transition parameter influences the basin entropy. Our paper ends with Section \ref{conc}, where we emphasize the main conclusions of this work.

\section{Properties of the dynamical system}
\label{mod}

Let us briefly recall the main properties of the dynamical systems under consideration. A dimensionless, barycentric rotating system of coordinates $Oxyz$, is considered for both cases. For the description of the planar motion of the test particle we choose a rotating reference frame, where the center of mass of the primaries coincides with its origin.

\begin{enumerate}
  \item \textbf{The pseudo-Newtonian circular restricted three-body problem}: Two primary bodies, $P_1$ and $P_2$, are located on the $Ox$ axis with masses $m_1 = \mu$ and $m_2 = 1 - \mu$, respectively, where $\mu = m_2/(m_1 + m_2) \leq 1/2$ is the mass parameter \cite{S67}. The centers of both primaries are located at $(x_1, 0, 0)$ and $(x_2, 0, 0)$, where $x_1 = - \mu$ and $x_2 = 1 - \mu$.

      According to Ref. \cite{DLCG17} the effective potential function of the pseudo-Newtonian circular restricted three-body problem, with only the first correction terms, is given by
      \begin{equation}
      \Omega_3(x,y,z) = \sum_{i=1}^{2} \frac{m_i}{r_{3i}} - \frac{\epsilon}{2 c^4} \sum_{i=1}^{2} \frac{m_i^3}{r_{3i}^3} + \frac{1}{2}\left(x^2 + y^2\right),
      \label{eff3}
      \end{equation}
      where of course
      \begin{equation}
      \begin{aligned}
      r_{31} &= \sqrt{\left(x - x_1\right)^2 + y^2 + z^2},\\
      r_{32} &= \sqrt{\left(x - x_2\right)^2 + y^2 + z^2},
      \label{dist3}
      \end{aligned}
      \end{equation}
      are the distances of the test particle from the two primary bodies.
  \item \textbf{The pseudo-Newtonian circular restricted four-body problem}: Three primary bodies, $P_1$, $P_2$, and $P_3$, with masses $m_1$, $m_2$, and $m_3$, respectively are located on the configuration $(x,y)$ plane, while their mutual distances remain constant and form an equilateral triangle. In the special case, where the three primaries have equal masses $(m_1 = m_2 = m_3 = 1/3)$ the coordinates of the centers are $(x_1,0,0)$, $(x_2,y_2,0)$, and $(x_3,y_3,0)$, where $x_1 = \sqrt{3}/3$, $x_2 = - \sqrt{3}/6$, $x_3 = x_2$, $y_2 = 1/2$, and $y_3 = - y_2$.

      In the same vein of the pseudo-Newtonian three-body problem, the effective potential function of the pseudo-Newtonian four-body problem (where again we consider only the first correction terms) is
      \begin{equation}
      \Omega_4(x,y,z) = \sum_{i=1}^{3} \frac{m_i}{r_{4i}} - \frac{\epsilon}{2 c^4} \sum_{i=1}^{3} \frac{m_i^3}{r_{4i}^3} + \frac{1}{2}\left(x^2 + y^2\right),
      \label{eff4}
      \end{equation}
      where once more
      \begin{equation}
      \begin{aligned}
      r_{41} &= \sqrt{\left(x - x_1\right)^2 + y^2 + z^2},\\
      r_{42} &= \sqrt{\left(x - x_2\right)^2 + \left(y - y_2\right)^2 + z^2},\\
      r_{43} &= \sqrt{\left(x - x_3\right)^2 + \left(y - y_3\right)^2 + z^2},
      \label{dist4}
      \end{aligned}
      \end{equation}
      are the distances of the third body from the three primaries. In \ref{appex} we prove that the equilateral triangular configuration of the system of three primary bodies is always unstable.
      \end{enumerate}

In both cases, the parameter $\epsilon$ acts as a transition parameter from classical Newtonian to pseudo-Newtonian dynamics, while its values lie in the interval $\epsilon \in [0, 1]$. Indeed, when $\epsilon = 0$ both effective potential functions, $\Omega_3(x,y,z)$ and $\Omega_4(x,y,z)$, are reduced to the corresponding versions of classical Newtonian dynamics. On the other hand, when $\epsilon = 1$ we have the case of the full pseudo-Newtonian problem.

We adopt a system of units in which the sum of the masses of the primaries, the angular velocity $(\omega)$, the speed of light $c$, the distance $R$ between the primaries as well as the gravitational constant $G$ are equal to unity.

The equations of motion describing a test particle (a body of a negligible mass $m$, with respect to the masses of the primaries) moving under the mutual gravitational attraction of the primaries read
\begin{equation}
\begin{aligned}
&\ddot{x} - 2 \dot{y} = \frac{\partial \Omega_i}{\partial x}, \\
&\ddot{y} + 2 \dot{x} = \frac{\partial \Omega_i}{\partial y}, \\
&\ddot{z} = \frac{\partial \Omega_i}{\partial z},
\label{eqmot}
\end{aligned}
\end{equation}
where $i = 3,4$.

The above system of differential equations admits only one integral of motion (also known as the Jacobi integral), which is
given by the following Hamiltonian
\begin{equation}
J(x,y,z,\dot{x},\dot{y},\dot{z}) = 2\Omega_i(x,y,z) - \left(\dot{x}^2 + \dot{y}^2 + \dot{z}^2 \right) = C,
\label{ham}
\end{equation}
where $\dot{x}$, $\dot{y}$, and $\dot{z}$ are the velocities, while $C$ is the numerical value of the Jacobi constant which is conserved.

\begin{figure}[!t]
\centering
\resizebox{0.75\hsize}{!}{\includegraphics{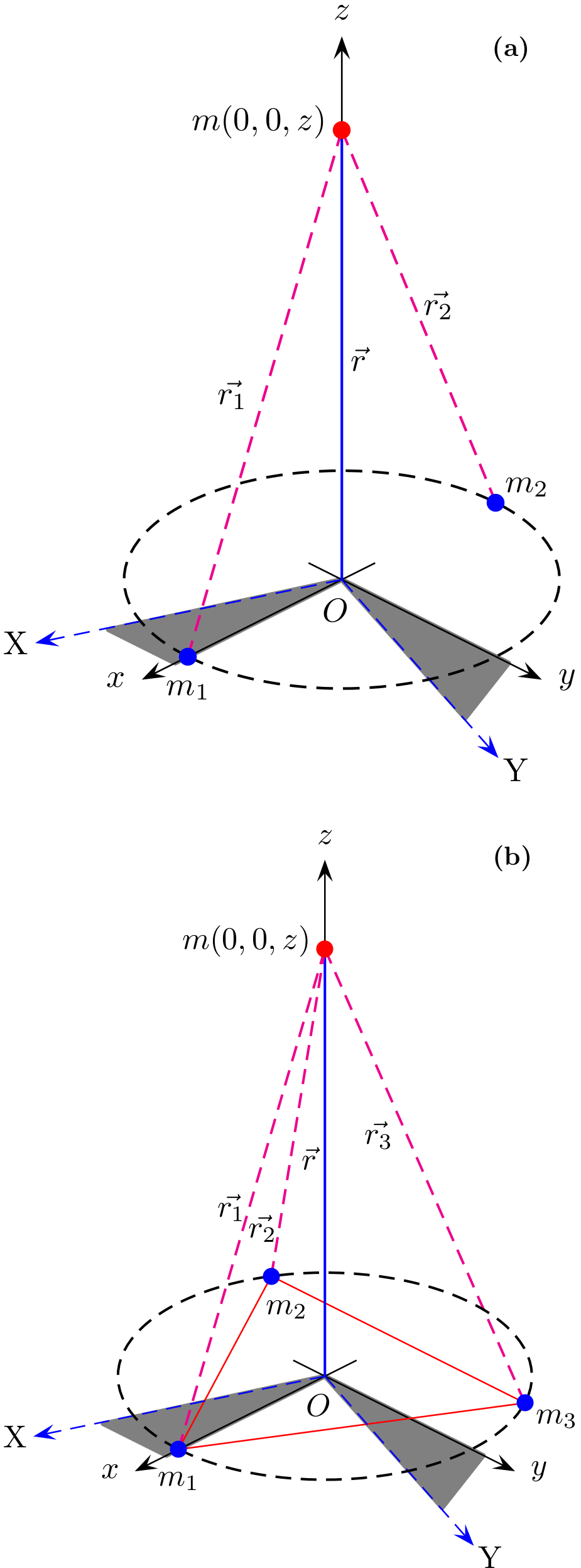}}
\caption{The configuration of the Sitnikov problem, where equally massed primary bodies move on symmetric circular orbits. (a-left): The case of two equally massed primaries $(m_1 = m_2 = 1/2)$ and (b-right): the case of three equally massed primaries $(m_1 = m_2 = m_3 = 1/3)$.}
\label{conf}
\end{figure}

The potential function of the circular Sitnikov problem can be obtained if we set equal masses to all primaries and $x = y = 0$. For the restricted three-body problem with $\mu = 1/2$ we have
\begin{equation}
\Omega_3(z) = \frac{1}{r_3} \left(1 - \frac{\epsilon}{8r_3^2}\right),
\label{potz3}
\end{equation}
where $r_3 = \sqrt{z^2 + 1/4}$. Similarly, the potential function of the circular Sitnikov pseudo-Newtonian four-body problem reads
\begin{equation}
\Omega_4(z) = \frac{1}{r_4} \left(1 - \frac{\epsilon}{18r_4^2}\right),
\label{potz4}
\end{equation}
where $r_4 = \sqrt{z^2 + 1/3}$. It is evident that Eqs. (\ref{potz3}) and (\ref{potz4}) describe the motion of a massless test particle which oscillates along a straight line which is perpendicular to the orbital configuration $(x,y)$ plane of the primary bodies with equal masses. In Fig. \ref{conf} we present the geometry of the Sitnikov problem, related to the pseudo-Newtonian circular three and four-body problems.

Consequently, the equations, regarding the motion of the test particle along the vertical $z$ axis, have the form
\begin{equation}
\begin{aligned}
\ddot{z} &= \frac{d}{d z}\left(\Omega_3(z)\right) = - \frac{z}{r_3^3} \left(1 - \frac{3\epsilon}{8r_3^2}\right), \\
\ddot{z} &= \frac{d}{d z}\left(\Omega_4(z)\right) = - \frac{z}{r_4^3} \left(1 - \frac{\epsilon}{6r_4^2}\right),
\label{eqmotz}
\end{aligned}
\end{equation}
while the corresponding Jacobi integral, for the vertical motion, becomes
\begin{equation}
J_z(z,\dot{z}) = 2 \Omega_i(z) - \dot{z}^2 = C_{z},
\label{hamz}
\end{equation}
with $i=3,4$.

\section{Parametric variation of the equilibrium points}
\label{param}

From now on, the $z$ coordinate is considered as a complex variable and it is denoted by $\mathz$, thus following the approach which was successfully used in Refs. \cite{DKMP12,Z17,ZSAS18,Z18a}. At this point, it should be explained that the transition to complex variables is imperative. This is because, as it was presented in Ref. \cite{D10}, all the impressive and beautiful fractal basin structures can be observed only on the complex plane.

In order to obtain the equilibrium points (roots) of the vertical motion, the right hand side of Eq. (\ref{eqmotz}) should be taken equal to zero as
\begin{equation}
\begin{aligned}
f_3(\mathz;\epsilon) &= - \frac{4\mathz \left(8\mathz^2 - 3\epsilon + 2\right)}{\left(1 + 4\mathz^2\right)^{5/2}} = 0, \\
f_4(\mathz;\epsilon) &= - \frac{3 \sqrt{3}\mathz \left(6\mathz^2 - \epsilon + 2\right)}{2\left(1 + 3\mathz^2\right)^{5/2}} = 0,
\label{fza0}
\end{aligned}
\end{equation}
which are reduced to
\begin{equation}
\begin{aligned}
&\mathz \left(8\mathz^2 - 3\epsilon + 2\right) = 0, \\
&\mathz \left(6\mathz^2 - \epsilon + 2\right) = 0.
\label{fza}
\end{aligned}
\end{equation}

\begin{figure*}[!t]
\centering
\resizebox{\hsize}{!}{\includegraphics{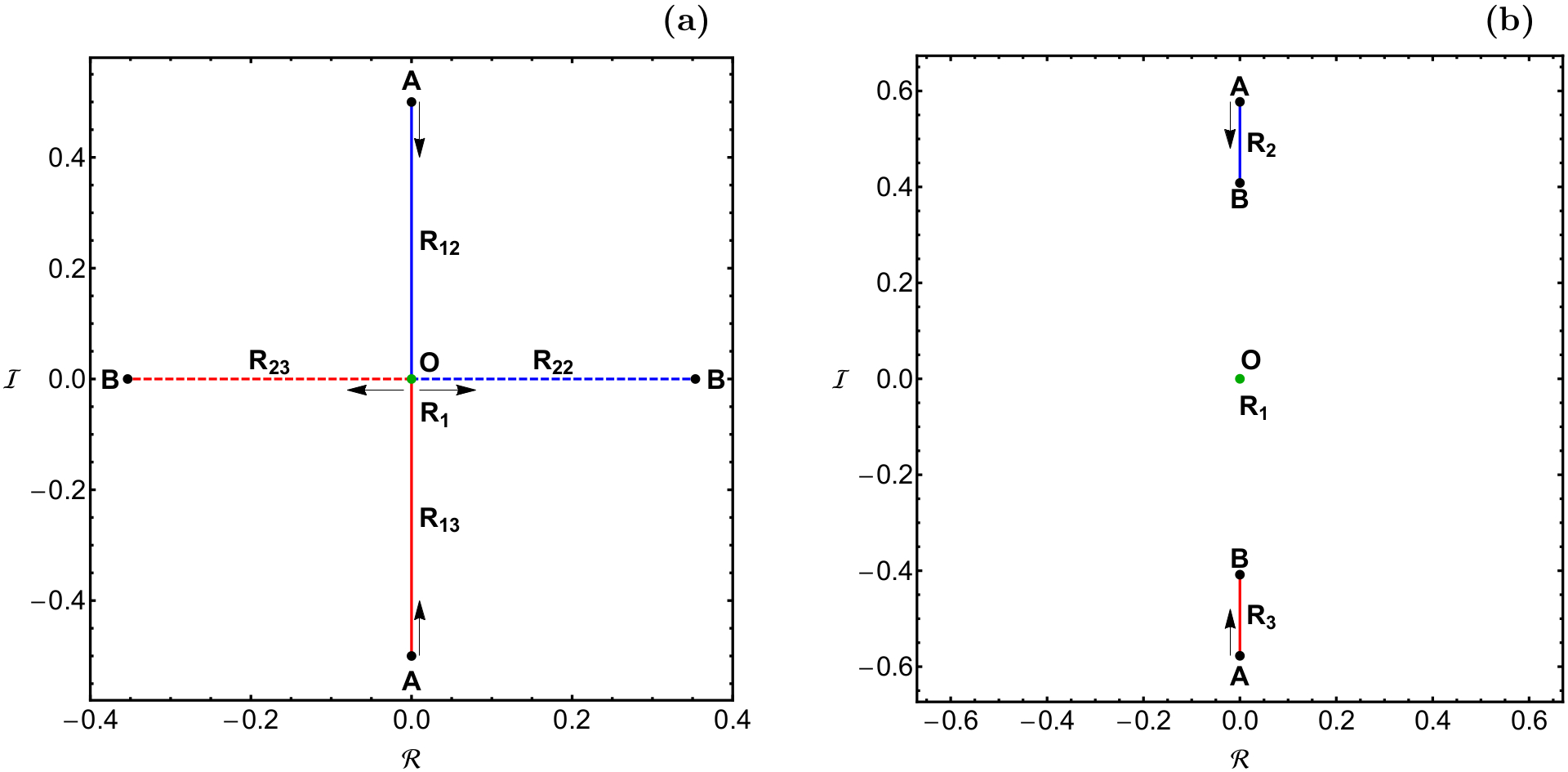}}
\caption{The space evolution of the roots on the complex plane, when $\epsilon \in [0,1]$. (a-left): The case when two primary bodies are present. When $\epsilon < 2/3$ we have the roots $R_{12}$ (solid blue) and $R_{13}$ (solid red) and when $\epsilon > 2/3$ we have the roots $R_{22}$ (dashed blue) and $R_{23}$ (dashed red). (b-right): The case when three primary bodies are present and two imaginary roots $R_2$ (blue) and $R_3$ (red) exist. The arrows indicate the movement direction of the roots, as the value of the transition parameter increases. The black dots (points A and B) correspond to the cases $\epsilon \to 0$ and $\epsilon = 1$, respectively, while the green point indicates the fixed central root $R_1$. (Color figure online).}
\label{evol}
\end{figure*}

Looking at Eqs. (\ref{fza}) we observe that the root $\mathz = 0$ is always present, regardless the value $\epsilon$ of the transition parameter. This root corresponds to the inner collinear equilibrium point $L_1$ of the circular restricted three and four-body problems. However since the left hand side of Eqs. (\ref{fza}) is a third order polynomial it means that there are two additional roots, given by
\begin{equation}
\mathz_i = \pm \frac{\sqrt{3\epsilon - 2}}{2\sqrt{2}}, \ \ \ i = 1,2,
\label{rts3}
\end{equation}
in the case of two primary bodies and
\begin{equation}
\mathz_i = \pm \frac{\sqrt{\epsilon - 2}}{\sqrt{6}}, \ \ \ i = 1,2,
\label{rts4}
\end{equation}
when three primaries are present.

The nature of these two roots strongly depends on the numerical value $\epsilon$ of the transition parameter. Our numerical analysis reveals that for the circular Sitnikov pseudo-Newtonian three-body problem, along with the $\mathz = 0$ root
\begin{itemize}
  \item When $\epsilon < -2/3$ there are two imaginary roots.
  \item When $\epsilon = 0$ or $\epsilon = 2/3$ only the root $\mathz = 0$ exists.
  \item When $\epsilon > 2/3$ there are two real roots.
\end{itemize}
while for the circular Sitnikov pseudo-Newtonian four-body problem applies that
\begin{itemize}
  \item When $\epsilon < -2$ there are two imaginary roots.
  \item When $\epsilon = 0$ or $\epsilon = 2$ only the root $\mathz = 0$ exists.
  \item When $\epsilon > 2$ there are two real roots.
\end{itemize}
However since the numerical value of the transition parameter $\epsilon$ cannot exceed 1, real roots are impossible for the circular Sitnikov pseudo-Newtonian four-body problem. It should be noted that $\epsilon = 1$ is the maximum allowed value of the transition parameter, corresponding to full pseudo-Newtonian dynamics, while all higher values $(\epsilon > 1)$ have no physical meaning. It is seen, that the value $\epsilon = 2/3$ is in fact a critical value of the transition parameter, since it determines the change on the nature of the two roots, in the case of the circular Sitnikov pseudo-Newtonian three-body problem.

\begin{figure*}[!t]
\centering
\resizebox{\hsize}{!}{\includegraphics{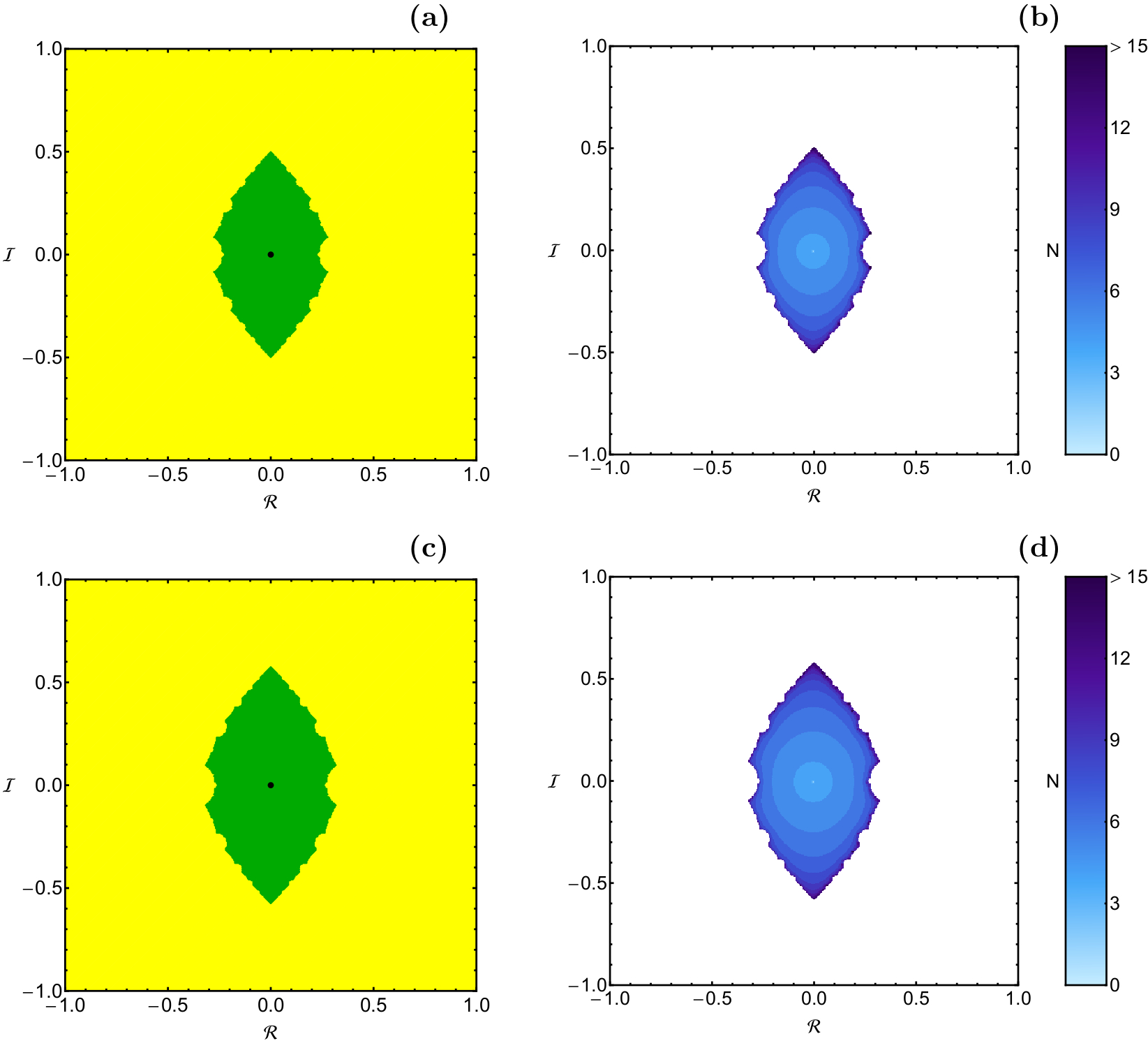}}
\caption{The Newton-Raphson basins of convergence on the complex plane, when $\epsilon = 0$, and only one root exists. (a-upper left): The case where two primaries are present and (c-lower left): The case where three primaries are present. The position of the root is indicated by a black dot. The color code is as follows: $R_1$ root (green); points tending to infinity (yellow); non-converging points (white). (Panels (b) and (d)): The distribution of the corresponding number $(N)$ of required iterations for obtaining the Newton-Raphson basins of convergence shown in panels (a) and (c), respectively. (Color figure online).}
\label{m0}
\end{figure*}

Useful information could be extracted from the parametric variation of the roots on the complex plane, as a function of the transition parameter. Fig. \ref{evol}(a-b) shows the parametric evolution of the roots on the complex plane, when $\epsilon \in [0,1]$, with $\mathcal{R} = Re[\mathz]$ and $\mathcal{I} = Im[\mathz]$. For the case where two primaries are present we see in panel (a) of Fig. \ref{evol} that as long as $\epsilon > 0$ two imaginary roots $R_{12}$ and $R_{13}$ appear. As the value of the transition parameter increases both imaginary roots tend to the origin and when $\epsilon = 2/3$ they collide with the central root $R_1$ and only the root $\mathz = 0$ survives thus having multiplicity 3. When $\epsilon > 2/3$ two real roots $R_{22}$ and $R_{23}$ emerge from the origin $O$ and they move away from each other, with increasing value of $\epsilon$. Finally, when $\epsilon = 1$ we the two real roots $\pm \sqrt{2}/4$ reach their maximum distance from the origin. When three primaries are present it is seen in panel (b) of Fig. \ref{evol} that the evolution of the imaginary roots $R_2$ and $R_3$ follow a similar evolution with respect to the first case (with roots $R_{12}$ and $R_{13}$) of two primaries. When $\epsilon = 1$ we have the imaginary roots $\pm i \sqrt{6}/6$ and their evolution is terminated, thus avoiding the collision with the origin.

\section{The Newton-Raphson basins of convergence}
\label{res}

\begin{figure*}[!t]
\centering
\resizebox{\hsize}{!}{\includegraphics{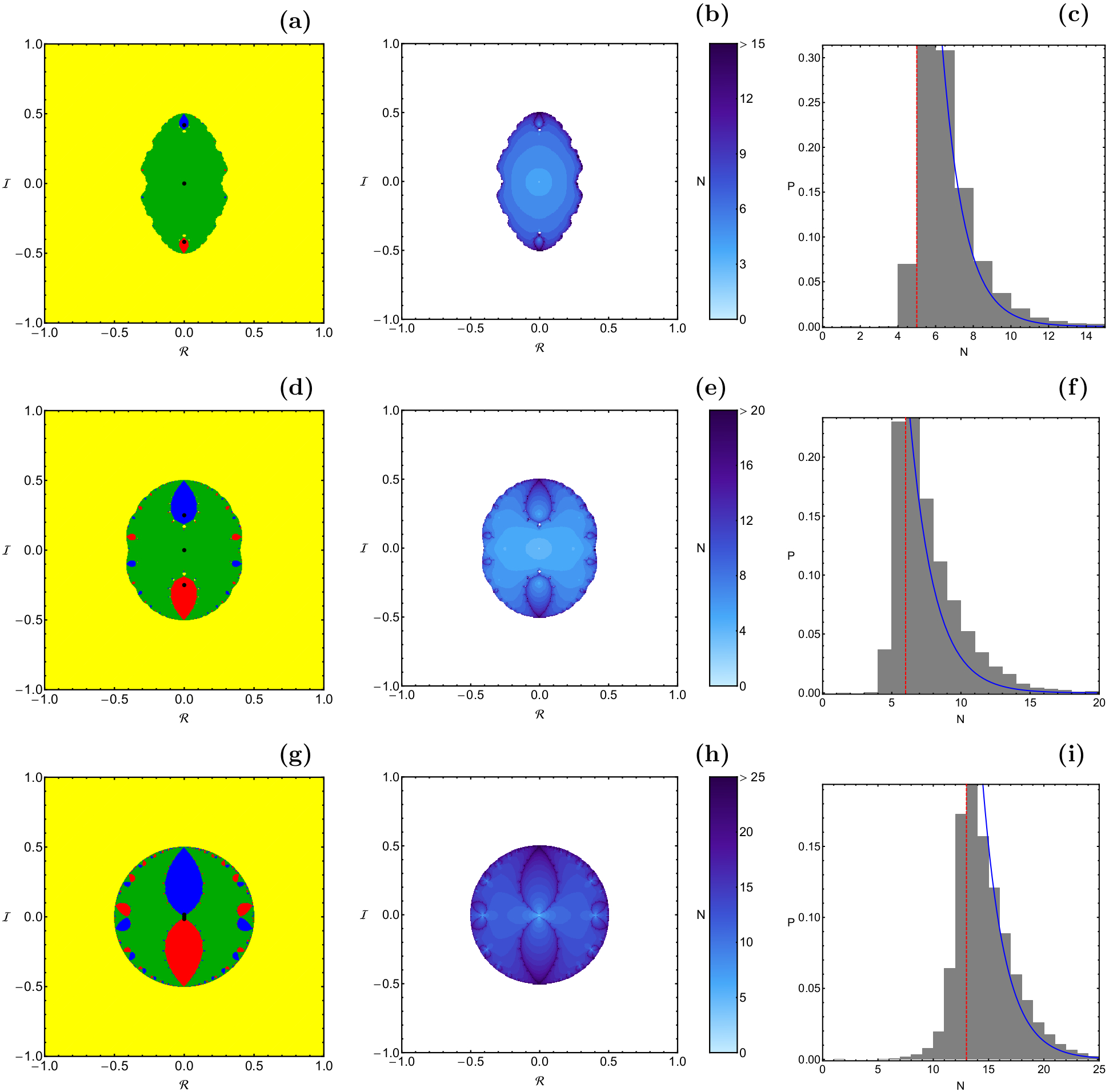}}
\caption{(First column): The Newton-Raphson basins of convergence on the complex plane for the first case, when two primaries are present and $0 < \epsilon < 2/3$. The color code, denoting the three roots, is as follows: $R_1$ (green); $R_{12}$ (blue); $R_{13}$ (red); tending to infinity (yellow); non-converging points (white). The positions of the three roots are indicated by black dots. (Second column): The distribution of the corresponding number $N$ of required iterations for obtaining the Newton-Raphson basins of convergence. The points tending to infinity as well as the non-converging points are shown in white. (Third column): The corresponding probability distribution of required iterations for obtaining the Newton-Raphson basins of convergence. The vertical dashed red line indicates, in each case, the most probable number $N^{*}$ of iterations. (First row): $\epsilon = 0.2$; (Second row): $\epsilon = 0.5$; (Third row): $\epsilon = 0.666$. (Color figure online).}
\label{m1}
\end{figure*}

\begin{figure*}[!t]
\centering
\resizebox{\hsize}{!}{\includegraphics{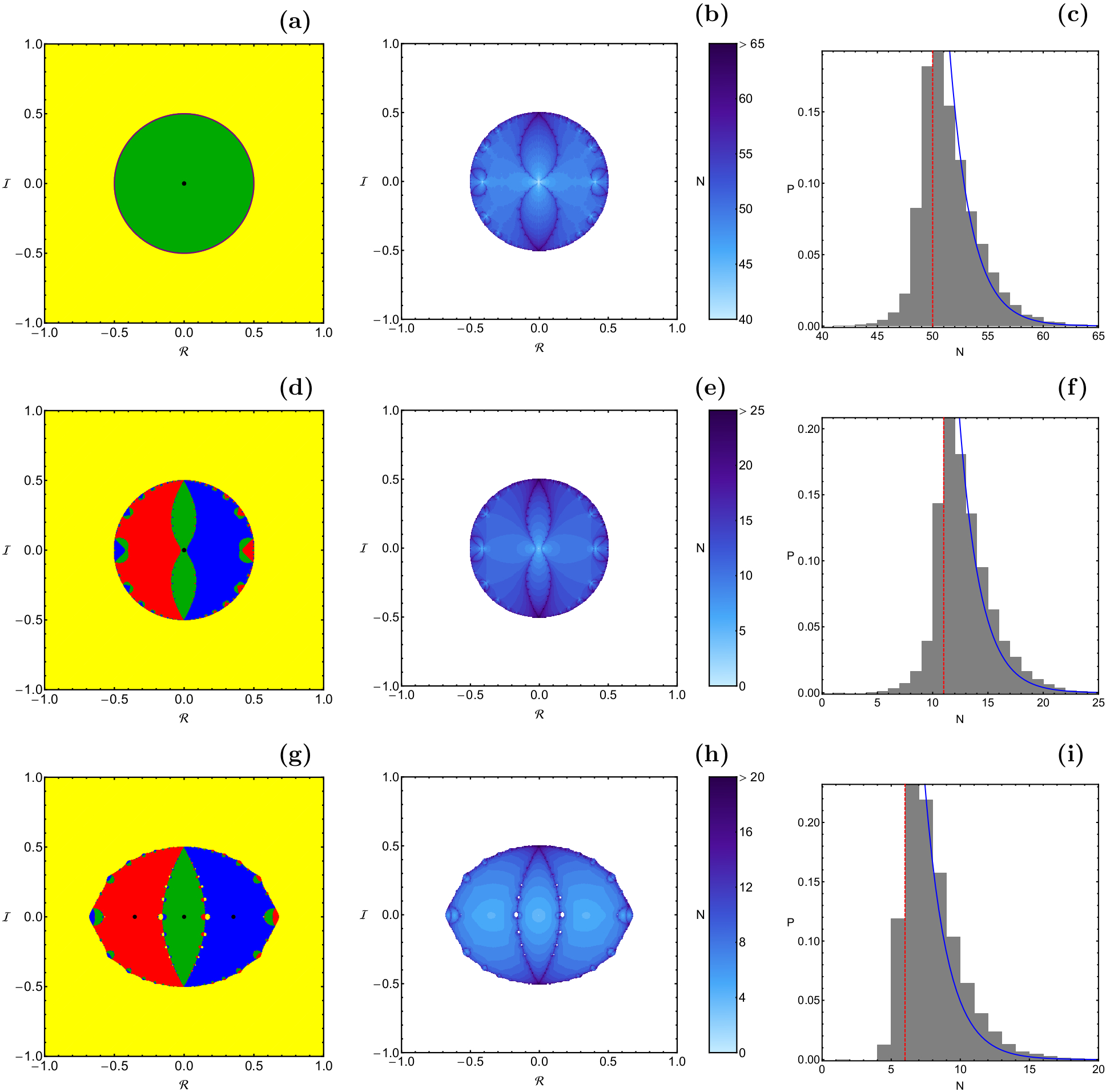}}
\caption{(First column): The Newton-Raphson basins of convergence on the complex plane for the second case, when two primaries are present and $2/3 \leq \epsilon \leq 1$. The color code, denoting the three roots, is as follows: $R_1$ (green); $R_{22}$ (blue); $R_{23}$ (red); tending to infinity (yellow); non-converging points (white). The positions of the three roots are indicated by black dots. (Second column): The distribution of the corresponding number $N$ of required iterations for obtaining the Newton-Raphson basins of convergence. The points tending to infinity as well as the non-converging points are shown in white. (Third column): The corresponding probability distribution of required iterations for obtaining the Newton-Raphson basins of convergence. The vertical dashed red line indicates, in each case, the most probable number $N^{*}$ of iterations. (First row): $\epsilon = 2/3$; (Second row): $\epsilon = 0.67$; (Third row): $\epsilon = 1$. (Color figure online).}
\label{m2}
\end{figure*}

\begin{figure*}[!t]
\centering
\resizebox{\hsize}{!}{\includegraphics{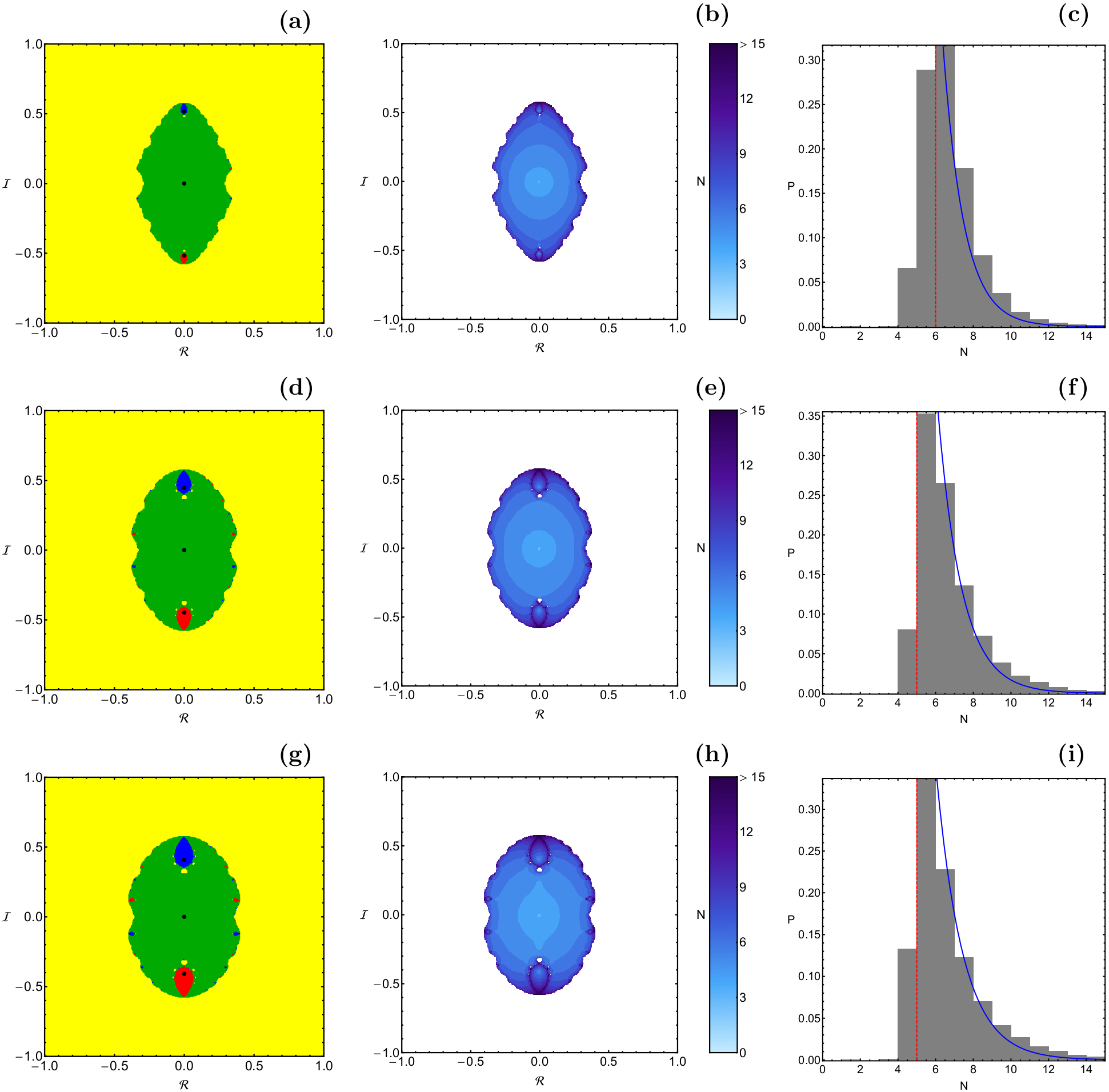}}
\caption{(First column): The Newton-Raphson basins of convergence on the complex plane for the third case, when three primaries are present and $0 < \epsilon \leq 1$. The color code, denoting the three roots, is as follows: $R_1$ (green); $R_2$ (blue); $R_3$ (red); tending to infinity (yellow); non-converging points (white). The positions of the three roots are indicated by black dots. (Second column): The distribution of the corresponding number $N$ of required iterations for obtaining the Newton-Raphson basins of convergence. The points tending to infinity as well as the non-converging points are shown in white. (Third column): The corresponding probability distribution of required iterations for obtaining the Newton-Raphson basins of convergence. The vertical dashed red line indicates, in each case, the most probable number $N^{*}$ of iterations. (First row): $\epsilon = 0.4$; (Second row): $\epsilon = 0.8$; (Third row): $\epsilon = 1$. (Color figure online).}
\label{m3}
\end{figure*}

\begin{figure*}[!t]
\centering
\resizebox{\hsize}{!}{\includegraphics{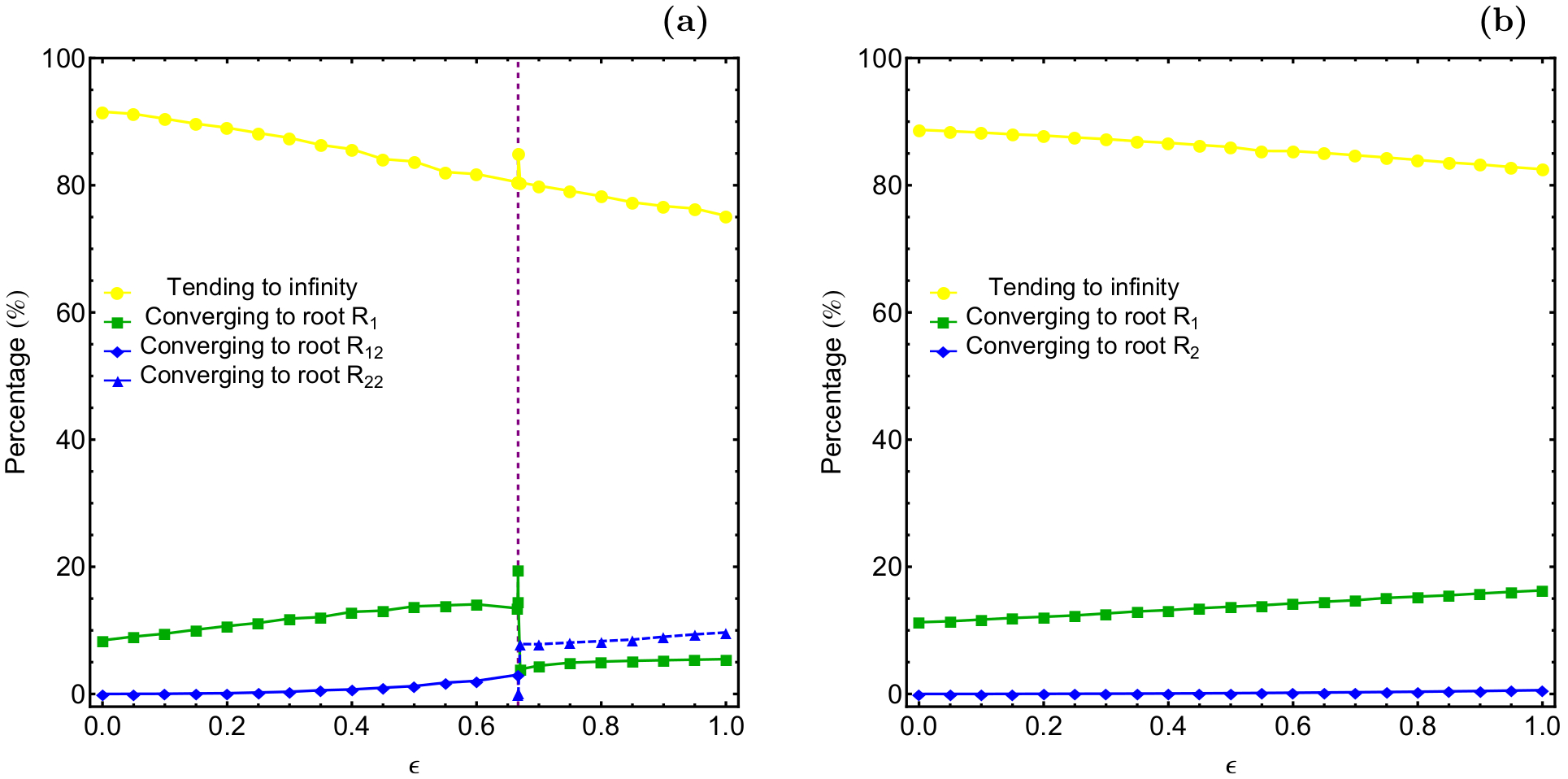}}
\caption{Evolution of the percentages of all types of initial conditions on the complex plane, as a function of the transition parameter $\epsilon$, when (a-left): two primaries are present and (b-right): three primaries are present. The vertical, dashed, purple line indicates the critical value $\epsilon = 2/3$. (Color figure online).}
\label{percs}
\end{figure*}

For obtaining the basins of convergence on the complex plane, associated with the roots of the system, we have to numerically solve the corresponding mono-parametric equation. The easiest way is to use the classical Newton-Raphson method of second order, through the iterative scheme
\begin{equation}
\mathz_{n+1} = \mathz_n - \frac{f_i(\mathz_n;\epsilon)}{f_i'(\mathz_n;\epsilon)}, \ \ \ i=3,4,
\label{nr0}
\end{equation}
where $\mathz_n$ is the value of the $\mathz$ at the $n$-th step of the iterative process. By plugging formulae (\ref{fza0}) in Eq. (\ref{nr0}) we obtain
\begin{equation}
\mathz_{n+1} = \frac{12\mathz^3 \left(8\mathz^2 - 5 \epsilon + 2 \right)}{8\mathz^2 \left(8\mathz^2 - 6 \epsilon + 1\right) + 3\epsilon - 2},
\label{nr3}
\end{equation}
for the case of two primary bodies and
\begin{equation}
\mathz_{n+1} = \frac{3\mathz^3 \left(18\mathz^2 - 5 \epsilon + 6 \right)}{6\mathz^2 \left(6\mathz^2 - 2 \epsilon + 1\right) + \epsilon - 2},
\label{nr4}
\end{equation}
when three primaries exist on the configuration plane.

The numerical algorithm for revealing the basins of convergence on the complex plane works as follows: The iterative procedure is activated by an initial condition of the form of a complex number $\mathz = a + ib$, with $\mathcal{R} = a$ and $\mathcal{I} = b$ and continues until a root is reached, with the predefined numerical accuracy. In our case the numerical accuracy is set to $10^{-15}$, for both real and imaginary parts. Using a dense uniform grid of $1024 \times 1024$ $(\mathcal{R},\mathcal{I})$ nodes, as initial conditions, we perform a double scan of the complex plane. During the iterative procedure we monitor the required number $N$ of iterations, along with the classification of the nodes. For our experiments the maximum allowed number of iterations is set to $N_{\rm max} = 500$.

If the iterative procedure leads to a root it means that the Newton-Raphson method converges for the particular initial condition. Here it should be clarified that the numerical method, in general terms, does not converge equally well for all the initial conditions on the complex plane. All the initial conditions which lead to the same root (final state of the iterative procedure) compose the Newton-Raphson basin of convergence/attraction (we will also use the terms attractive regions/domains). Furthermore, we would like to emphasize that the Newton-Raphson basins of convergence should not be mistaken with the basins of attraction which are encountered in dissipative systems.

The Newton-Raphson basins of convergence when $\epsilon = 0$ are presented in panels (a), for the case of two primaries, and (c), for the case of three primaries, of Fig. \ref{m0}. We see that the converging initial conditions are mainly located near the center, while they form a rhomboidal fractal shape. On the other hand, the vast majority of the complex plane is covered by a unified sea (yellow region) of initial conditions for which the Newton-Raphson iterative scheme tends very quickly to extremely large complex numbers. This numerical behaviour implies that for these initial conditions tend, theoretically, to infinity. In both cases (two and three primaries) the overall geometry of the converging regions are qualitatively very similar. This can be explained, in a way, if we examine the form of the corresponding iterative schemes. For the case of two primary bodies the Newton-Raphson iterative scheme with $\epsilon = 0$ takes the form
\begin{equation}
\mathz_{n+1} = \frac{12\mathz^3}{8\mathz^2 - 1},
\label{iter3}
\end{equation}
while for the case of three primaries we have that
\begin{equation}
\mathz_{n+1} = \frac{9\mathz^3}{6\mathz^2 - 1}.
\label{iter4}
\end{equation}
We observe that the complex functions entering both iterative schemes are completely identical and only the numerical coefficients are different. This should be the main reason of the similar patterns observed in panels (a) and (c) of Fig. \ref{m0}. In panels (b) and (d) of the same figure the distribution of the corresponding number $(N)$ of iterations required for obtaining the desired accuracy is given using tones of blue.

In the following subsections we will determine how the transition parameter $\epsilon$ affects the structure of the Newton-Raphson basins of convergence in the circular Sitnikov problem with two and three primary bodies. For the classification of the nodes on the complex plane we will use color-coded basin diagrams, in which each pixel is assigned a different color, according to the final state (root) of the corresponding initial condition.

\subsection{Case I: Two primary bodies present and $0 < \epsilon < 2/3$}
\label{ss1}

We begin with the first case where two primary bodies are present on the configuration $(x,y)$ plane, while the equation $f_3(\mathz;\epsilon) = 0$ has, apart from the $\mathz = 0$ root, two imaginary roots. The Newton-Raphson basins of convergence on the complex plane, for three values of the transition parameter, are illustrated in the first column of Fig. \ref{m1}. It is seen that in all cases the area of all the types of the basins of convergence is finite. On the contrary, outside the convergence regions the vast majority of the complex plane is covered by initial conditions which do not converge to any of the three roots (yellow regions) but they tend asymptotically to infinity.

In the second column of Fig. \ref{m1} we present the corresponding number $N$ of iterations, using tones of blue, while the corresponding probability distribution of the required iterations is given in the third column of the same figure. The definition of the probability $P$ is the following: if $N_0$ complex initial conditions $(\mathcal{R},\mathcal{I})$ converge, after $N$ iterations, to one of the roots then $P = N_0/N_t$, where $N_t$ is the total number of nodes in every basin diagram. Moreover, in all plots the tails of the histograms extend so as to cover 97\% of the corresponding distributions of iterations. The vertical, red, dashed lines in the probability histograms denote the most probable number $N^{*}$ of iterations. The blue lines in the histograms of Fig. \ref{m1} indicate the best fit to the right-hand side $N > N^{*}$ of them (more details are given in subsection \ref{geno}).

We see that as soon as $\epsilon > 0$ several lobes emerge at the boundaries of the convergence region corresponding to the central root $R_1$. With increasing value of the transition parameter the most important changes which occur on the complex plane are the following:
\begin{itemize}
  \item The area of the basins of convergence corresponding to all three roots increase, while at the same time the overall shape of the convergence region changes from rhomboidal to spherical.
  \item The probability distribution of the required iterations moves to the right, which implies that additional iterations are required. Indeed, we observe that the most probable number $N^{*}$ of iterations increases from 6, when $\epsilon = 0.2$ to 13 when $\epsilon = 0.666$.
  \item In panel (h), where $\epsilon = 0.666$ it is seen that throughout the convergence regions the corresponding iterations are significantly higher with respect to panel (b), where $\epsilon = 0.2$. We suspect that this behaviour is directly related with the fact that we approaching the critical value $\epsilon = 2/3$.
\end{itemize}

\subsection{Case II: Two primary bodies present and $2/3 \leq \epsilon \leq 1$}
\label{ss2}

The next case under consideration involves the scenario where there are two real roots, along with the $\mathz = 0$ root. In the first column of Fig. \ref{m2} we present the Newton-Raphson basins of convergence for three values of the transition parameter. In panel (a) of Fig. \ref{m2}, which corresponds to the critical value of $\epsilon$ one can see that the convergence region corresponding to root $R_1$ is a perfect circle with radius 0.5. However, in panel (b) of the same figure one can observe that the initial conditions which form the circular region produce interesting inner structures, depending on the number of the required iterations.

As we proceed to higher values of the transition parameter the main changes, regarding the geometry of the convergence areas, are the following:
\begin{itemize}
  \item The extent of the basins of convergence of all three roots increases, while the geometry of the overall converging region alters from fractal spherical to elliptic (oval). As long as $\epsilon > 2/3$ two main lobes appear, corresponding to basins of convergence of the central roots, while these are lobes gradually unified thus producing a singe region.
  \item The probability distribution of the required iterations moves to the left, which implies that less iterations are required. Indeed, it is observed that the most probable number $N^{*}$ of iterations decreases from 50, when $\epsilon = 2/3$ to 6 when $\epsilon = 1$.
  \item The basin boundaries become more noisy, which means that the fractality of the several basin boundaries on the complex plane increases.
\end{itemize}

\subsection{Case III: Three primary bodies present}
\label{ss3}

We close with the last case where three primary bodies circulate on the configuration $(x,y)$ plane, while the equation $f_4(\mathz;\epsilon) = 0$ has two imaginary roots, along with the $R_1$ root $\mathz = 0$. The evolution of the geometry of the Newton-Raphson basins of convergence is depicted in the first column of Fig. \ref{m3}, where we present three basin diagrams for three values of the transition parameter. It becomes evident that this is the least interesting case, from the dynamical point of view.

Indeed, with increasing value of $\epsilon$ the overall structure of the complex plane hardly changes and the most important phenomena which take place are the following:
\begin{itemize}
  \item The area of the basin of convergence, corresponding to the central root $R_1$, increases, while at the same time a smaller increase on the extent of the other types of converging regions (corresponding to roots $R_2$ and $R_3$) is observed.
  \item Apart from the central lobes, corresponding to roots $R_2$ and $R_3$, several smaller convergence regions emerge at the boundaries of the main basin of convergence, corresponding to root $R_1$.
  \item The most probable number $N^{*}$ of iterations displays a minor reduce from 6, when $\epsilon = 0.4$, to 5, when $\epsilon = 1$.
\end{itemize}

\subsection{An overview analysis}
\label{geno}

Since the basin diagrams, presented earlier in subsections \ref{ss1}, \ref{ss2}, and \ref{ss3}, have a fixed and equal size we could monitor the evolution of the percentages of the different types of initial conditions, as a function of the transition parameter $\epsilon$. Such diagrams are presented in Fig. \ref{percs}(a-b). It is seen that in both cases, regarding the number (two or three) of the primary bodies, the percentage of initial conditions for which the Newton-Raphson iterative scheme tends to infinity is constantly reduced. On the other hand, the rates of all the types of converging initial conditions are increased, following an almost linear growth. In Panel (a) of Fig. \ref{percs} we see that the evolution of all types of the initial conditions is very smooth, except near the critical value $\epsilon = 2/3$, where the dynamical properties of the system change.

Additional interesting information could be extracted from the probability distributions of iterations presented in the third row of the basin diagrams. More precisely, we could try to obtain the best fit for the tails of the probability histograms. Using the term ``tails" we refer to the right-hand side of the histograms, where $N > N^{*}$, where of course $N^{*}$ is the most probable value. The ideal choice for fitting the tails of the histograms is the Laplace distribution, due to the fact that this particular type of distribution is commonly used in systems where transient chaos is present (see e.g., Refs. \cite{ML01,SASL06,SS08}). Our analysis indicates that the Laplace distribution is, in the vast majority of the cases, the best fit for the corresponding data.

For the Laplace distribution the corresponding probability density function (PDF) is defined as
\begin{equation}
P(N | a,b) = \frac{1}{2b}
 \begin{cases}
      \exp\left(- \frac{a - N}{b} \right), & \text{if } N < a \\
      \exp\left(- \frac{N - a}{b} \right), & \text{if } N \geq a
 \end{cases}.
\label{pdf}
\end{equation}
It is seen that two quantities enter the function $P(N)$. The first one $a$ is widely known as the location parameter, while $b$ is of course the diversity. Note that in our case, only the $N \geq a$ part of the distribution function is needed for fitting the tails of the histograms.

The aggregated results regarding the numerical values of the location parameter as well as the diversity, for all the studied cases, are presented in Table \ref{tab1}. It is very interesting to note that the most probable number of iterations is, in most of the cases, very close to the location parameter $a$, while in some of them both these quantities completely coincide, thus indicating the ideal choice of the Laplace distribution.

\begin{table}[!ht]
\begin{center}
   \caption{Aggregated results regarding the values of the most probable number of iterations $N^{*}$, the location parameter $a$ and the diversity $b$, regarding the basin diagrams shown earlier.}
   \label{tab1}
   \setlength{\tabcolsep}{10pt}
   \begin{tabular}{@{}lrrrr}
      \hline
      Figure & $\epsilon$ & $N^{*}$ & $a$ & $b$ \\
      \hline
      \ref{m0}a &     0 &  6 & $N^{*}$     & 1.15 \\
      \hline
      \ref{m1}c &   0.2 &  5 & $N^{*} + 1$ & 1.18 \\
      \ref{m1}f &   0.5 &  6 & $N^{*}$     & 1.80 \\
      \ref{m1}i & 0.666 & 13 & $N^{*} + 1$ & 2.03 \\
      \hline
      \ref{m2}c &   2/3 & 50 & $N^{*} + 1$ & 1.97 \\
      \ref{m2}f &  0.67 & 11 & $N^{*} + 1$ & 1.92 \\
      \ref{m2}i &     1 &  6 & $N^{*} + 1$ & 1.65 \\
      \hline
      \hline
      \ref{m0}c &     0 &  6 & $N^{*}$     & 1.15 \\
      \hline
      \ref{m3}c &   0.4 &  6 & $N^{*}$     & 1.11 \\
      \ref{m3}f &   0.8 &  5 & $N^{*} + 1$ & 1.27 \\
      \ref{m3}i &     1 &  5 & $N^{*} + 1$ & 1.42 \\
      \hline
   \end{tabular}
\end{center}
\end{table}

\section{Parametric evolution of the basin entropy}
\label{bas}

\begin{figure}[!t]
\centering
\resizebox{\hsize}{!}{\includegraphics{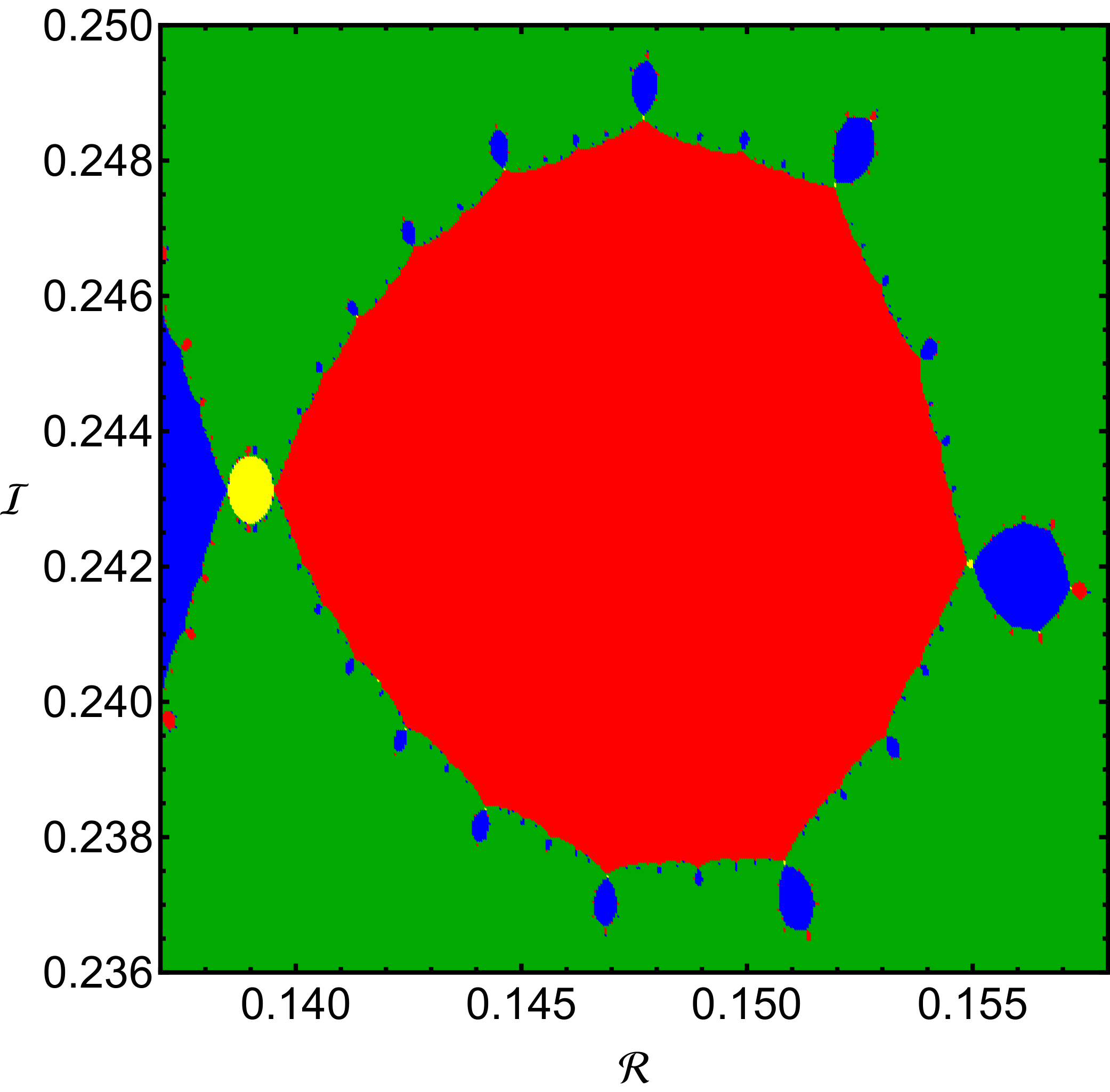}}
\caption{A magnification of a local area showing clearly the fractal basin boundaries. The local area corresponds to panel (g) of the basin diagram of Fig. \ref{m1}, where $\epsilon = 0.666$. The color code is the same as in Fig. \ref{m1}. (Color figure online).}
\label{zm}
\end{figure}

\begin{figure*}[!t]
\centering
\resizebox{\hsize}{!}{\includegraphics{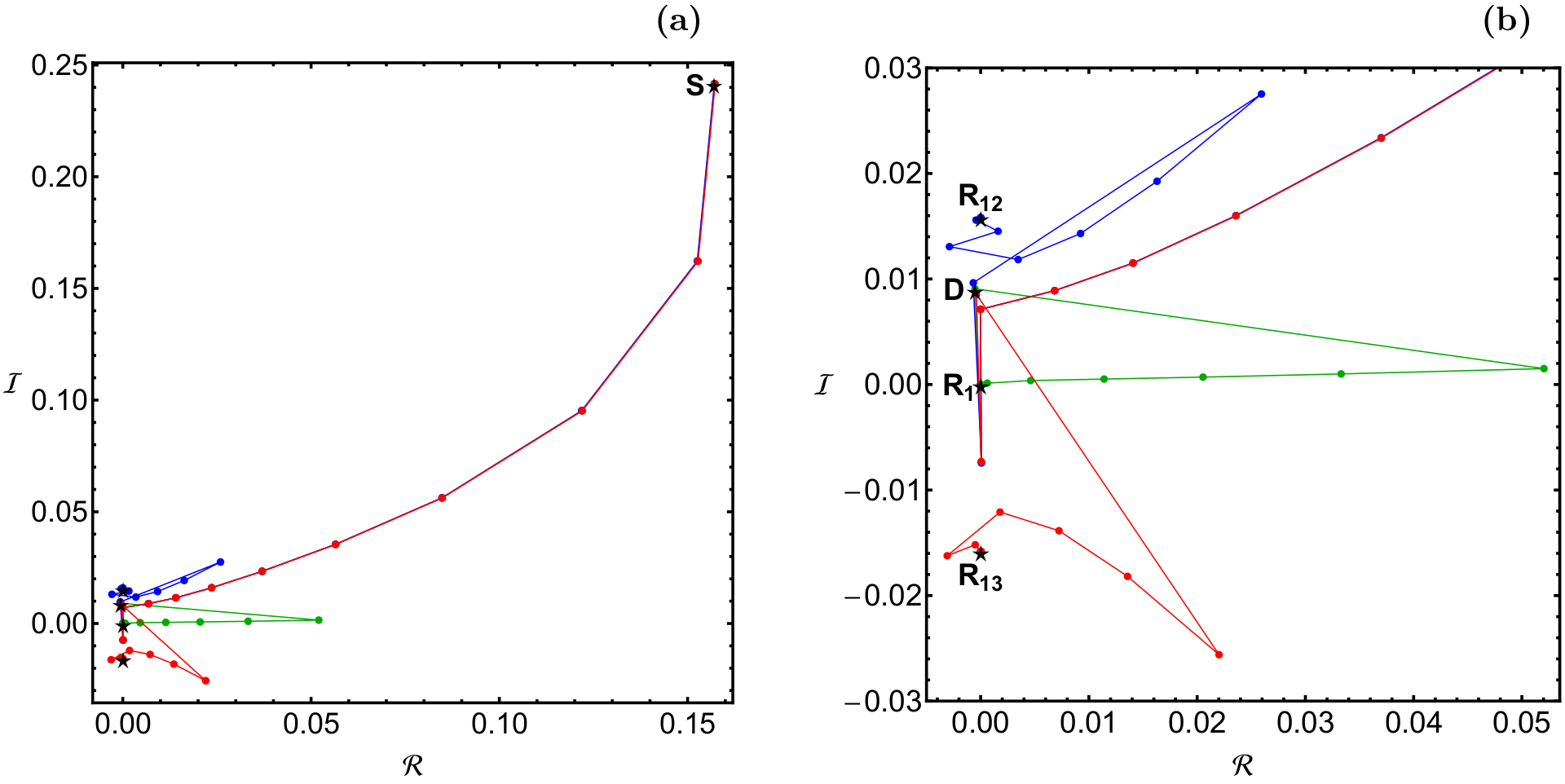}}
\caption{(a-left): A characteristic example of the consecutive steps that are followed by the Newton-Raphson iterator and the corresponding crooked path-lines of three nodes with complex initial conditions: (0.1572,0.2416) (green); (0.1571,0.2416) (blue); (0.1573,0.2416) (red). (b-right): A magnification of panel (a). Letter \textbf{S} denotes the starting points, while letter \textbf{D} indicates the point where the Newton-Raphson iterator starts to diverge for the three initial conditions, thus leading to three different final states. The positions of the three roots $R_1$, $R_{12}$, and $R_{13}$ are pinpointed by five-pointed stars. (Color figure online).}
\label{nr}
\end{figure*}

In the basin diagrams of the previous section we observed the presence of highly fractal regions, mainly located near the vicinity of the basin boundaries. By using the term fractal we simply imply that the particular areas display a fractal-like geometry, however without computing the corresponding fractal dimension as in Refs. \cite{AVS01,AVS09}. In Fig. \ref{zm} we present a magnification of a local area corresponding to the basin diagram shown earlier in panel (g) of Fig. \ref{m1}, where $\epsilon = 0.666$. Here the fractal basin boundaries are much more clear and distinct. It is known that the final state (root) of initial conditions inside these fractal areas is highly sensitive. Specifically, even the slightest change of the initial conditions automatically leads to a completely different root, which is a classical indication of chaos. Therefore, for the initial conditions located close to the basin boundaries it is almost impossible to predict their final states (roots).

To demonstrate the sensitivity of the fractal regions we chose three initial conditions inside a fractal region of Fig. \ref{zm}. All three initial conditions have the same imaginary part (0.2416), while their real parts differ only at the fourth decimal figure, which implies that the initial conditions are extremely close with each other. In panel (a) of Fig. \ref{nr} we depict the crooked path line created by the successive approximations-points that are followed by the Newton-Raphson iterator. We see that for the first eleven iterations the Newton-Raphson iterator follows almost identical paths for all three initial conditions. However after the eleventh iteration the three paths start to diverge, thus leading to three different final states (roots $R_1$, $R_{12}$, and $R_{13}$). Similar behavior applies for all the nodes with initial conditions inside the fractal areas of the basin diagrams.

In the previous section for describing the degree of fractality of the basin diagrams on the complex plane we used only qualitative arguments. In order to enrich our analysis we must also present quantitative information, regarding the fractal geometry of the basins of convergence. For this purpose we decided to use the basin entropy which was recently introduced in \cite{DWGGS16} and measures the degree of the basin fractality (or unpredictability). This new tool examines the topological properties of the convergence regions and provides quantitative results about their fractality.

\begin{figure*}[!t]
\centering
\resizebox{\hsize}{!}{\includegraphics{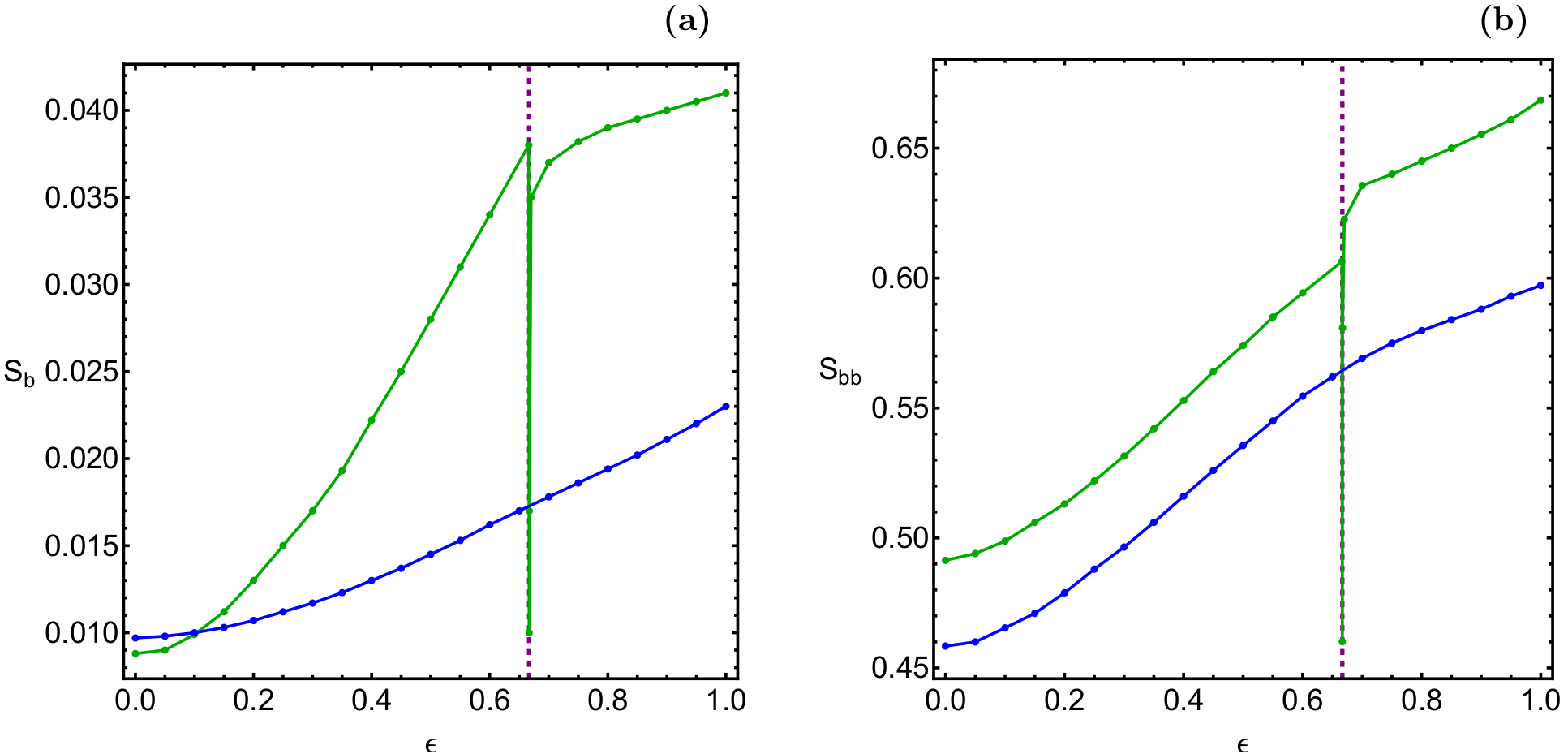}}
\caption{Evolution of (a-left): the basin entropy $S_b$ and (b-right): the boundary basin entropy $S_{bb}$, of the complex plane, as a function of the transition parameter $\epsilon$, when two (green) and three (blue) primary bodies are present on the configuration plane. The vertical, dashed, purple line indicates the critical value $\epsilon = 2/3$. (Color figure online).}
\label{frac}
\end{figure*}

The algorithm which describes how the basin entropy works is the following: Let us assume that we define a certain region $R = [-1,1] \times [-1,1]$ on the complex plane which we later divide into a rectangular grid of $N$ squares boxes (cells). Inside each of these cells there might exist between 1 and $N(A)$ attractors (roots in our case). Then we denote as $P_{i,j}$ the probability that inside a square box $i$ the attractor is $j$. Obviously, the initial conditions (nodes), inside each box, are completely independent. Therefore, the Gibbs entropy of each square cell is given by
\begin{equation}
S_{i} = \sum_{j=1}^{m_{i}}P_{i,j}\log_{10}\left(\frac{1}{P_{i,j}}\right),
\end{equation}
where $m_{i} \in [1,N_{A}]$ is the total number of the attractors (roots) that exist inside each square box $i$.

Adding all the individual entropies of the $N$ square cells of the rectangular grid on the complex plane we obtain the total entropy of the region $R$ as
\begin{equation}
S = \sum_{i=1}^{N} S_{i}.
\end{equation}
Consequently, the total entropy, associated with the entire amount of $N$ cells, is called basin entropy and it can be calculated as
\begin{equation}
S_{b} = \frac{1}{N}\sum_{i=1}^{N}\sum_{j=1}^{m_{i}}P_{i,j}\log_{10}\left(\frac{1}{P_{i,j}}\right).
\end{equation}

Using the value $\varepsilon = 0.005$, as suggested in Ref. \cite{DWGGS16}, we calculated the numerical value of the basin entropy $S_b$ of the complex plane, for several values of the transition parameter $\epsilon$. At this point, it should be emphasized that the initial conditions, on the complex plane, for which the Newton-Raphson iterative scheme fails to converge, were counted as an additional type of basin, which coexist along with the regular basins of convergence, associated with the equilibrium points of the system. Panel (a) of Fig. \ref{frac} illustrates the parametric evolution of the basins entropy, as a function of the transition parameter. Here it should be noted that for this diagram we used results not only from the cases, of Figs. \ref{m1}, \ref{m2}, and \ref{m3}, but also from additional levels of the transition parameter $\epsilon$.

One may observe that as the value of the transition parameter increases the basin entropy also increases for both cases (two and three primary bodies). Looking the diagram of Fig. \ref{frac}a we can identify two phenomena which should be emphasized: (i) at the critical value of the transition parameter the basin entropy displays a sudden as well as huge drop, almost to zero however, as soon as $\epsilon > 2/3$ it climbs up again and (ii) at the full pseudo-Newtonian case, where $\epsilon = 1$, the basin entropy, corresponding to the Sitnikov problem with two primary bodies, has almost twice the value of the case of the Sitnikov problem with three primaries.

Despite the thorough study of the basins of convergence, their fractal boundaries are only visible in magnifications of local areas of the basin diagrams (see e.g., Fig. \ref{zm}). In order to obtain quantitative results regarding the fractal nature of the basins we computed the boundary basin entropy \cite{DWGGS16}
\begin{equation}
S_{bb} = \frac{S}{N_b}
\label{sbb}
\end{equation}
where $N_b$ is the number of boxes containing more than one attractor.

According to the so-called ``log 2 criterion", the basin boundaries are not smooth (fractal) if $S_{bb} > \log 2$ (the converse is not true). In panel (b) of Fig. \ref{frac} we demonstrate the evolution of the boundary basin entropy $S_{bb}$, as a function of the transition parameter $\epsilon$. It is evident that in both cases (systems of two and three primary bodies) it is always $S_{bb} > \log 2$ which implies that the basin boundaries are always fractal in both systems under consideration.

\section{Concluding remarks}
\label{conc}

The convergence properties of the pseudo-Newtonian circular Sitnikov problem of three and four primary bodies have been numerically investigated. Using the classical, yet very effective, Newton-Raphson iterative scheme, we managed to reveal the basins of convergence on the complex plane. Additionally, we successfully determined the influence of the transition parameter $\epsilon$ on the roots as well as on several important quantities of the system.

As far as we know, there are no previous related numerical studies on the convergence properties of this particular dynamical system. Therefore, all the contained outcomes are novel and add to our existing knowledge on the basins of convergence in dynamical systems.

The following list contains the most important results of our numerical investigation:
\begin{enumerate}
  \item Imaginary roots are possible for both cases, regarding the number of the primary bodies (two and three). Real roots on the other hand, are possible only in the case of the Sitnikov problem corresponding to the pseudo-Newtonian circular restricted three-body problem.
  \item It was found that all the basins of convergence, corresponding to all three roots, have finite area, regardless the value of the transition parameter. Our numerical analysis indicates that the vast majority of the complex plane is covered by initial conditions which do not converge to any of the three roots. Furthermore, additional computations revealed that for all these initial conditions the Newton-Raphson iterator lead very fast to extremely large complex numbers (either real or imaginary), which implies that for these initial conditions the numerical method converges to the infinity.
  \item It should be emphasized that our classification of the initial conditions on the complex plane did not report any false-converging nodes. It should be explained that by the term ``false-converging" nodes we refer to initial conditions for which the iterative scheme leads (for $N < N_{\rm max}$) to final states which are not roots of the system, thus displaying a false convergence.
  \item Near the critical value of the transition parameter we identified several types of converging areas for which the corresponding number of required iterations is relatively high, with respect to near by basins of other roots. We suspect that this phenomenon is inextricably linked with the fact that near the critical point the dynamics of the system, such as the total number of the equilibrium points (roots), drastically changes.
  \item It was observed that the basin entropy of the complex plane is highly influenced by the transition parameter. More precisely, the highest value of $S_b$ is exhibited when $\epsilon \to 1$ (post-Newtonian dynamics), while on the other hand the basin entropy tends to zero when $\epsilon \to 0$ (classical Newtonian dynamics).
\end{enumerate}

For classifying the nodes on the complex plane we used a double precision \verb!FORTRAN 77! code \cite{PTVF92}. The required CPU time, per grid of initial conditions, was less then 5 minutes, using an Intel$^{\circledR}$ Quad-Core\textsuperscript{TM} i7 2.4 GHz PC. The latest version 11.3 of the software Mathematica$^{\circledR}$ \cite{W03} was used for developing all the graphical illustration of the paper.

In this article we used the simplest iterative method (the Newton-Raphson) for reveling the basins of convergence on the complex plane. An undeniably challenging task for a future work would be to use other iterative schemes (of order $n > 2$) and determine the corresponding similarities and differences on the convergence properties.

\section*{Acknowledgments}

The authors would like to express their warmest thanks to the two anonymous referees for the careful reading of the manuscript and for all the apt suggestions and comments which allowed us to improve both the quality and the clarity of the paper.

\begin{appendix}

\section{Stability of the equilateral triangular configuration of the four-body problem}
\label{appex}

According to \cite{R75} the stability of the equilateral triangular configuration of the four-body system strongly depends on the masses of the three primaries, as well as on the type of forces between the three main bodies. In particular, the system is stable only if
\begin{equation}
\frac{\left(m_1 + m_2 + m_3 \right)^2}{m_1 m_2 + m_1 m_3 + m_2 m_3} > 3 \left(\frac{1 + \kappa}{3 - \kappa}\right)^2,
\label{crit}
\end{equation}
where $\kappa$ is the power of the attraction law between the primaries.

For the case where all three primary bodies have equal masses (that is for $m_1 = m_2 = m_3 = 1/3$), it can be easily derived, from the above criterion, that the equilateral triangular configuration of the tree primaries is unstable not only in the classical Newtonian problem (where $\kappa = 2$) but also in the pseudo-Newtonian problem (where $\kappa = 4$).

\end{appendix}

\end{document}